\documentclass[prl,aps,reprint,longbibliography]{revtex4-1}

\usepackage{amsmath}
\usepackage{amssymb}
\usepackage{bm}
\usepackage{graphicx}
\usepackage{subfigure}
\usepackage{multirow}

\usepackage{xcolor}
\usepackage{hyperref}

\usepackage{natbib}

\usepackage{booktabs}

\begin{document}

\title{Minimum Long-Loop Feedback Vertex Set and Network Dismantling}
\author{Tianyi Li$^{1}$}
\thanks{tianyil@mit.edu}
\author{Pan Zhang$^{2,3,4}$}
\thanks{panzhang@itp.ac.cn}
\author{Hai-Jun Zhou$^{2,5,6}$}
\thanks{zhouhj@itp.ac.cn}

\affiliation{$^1$System Dynamics Group, Sloan School of Management, Massachusetts Institute of Technology \\
$^2$CAS Key Laboratory for Theoretical Physics, Institute of Theoretical Physics, Chinese Academy of Sciences, Beijing 100190, China \\
$^3$School of Fundamental Physics and Mathematical Sciences, Hangzhou Institute for Advanced Study, UCAS, Hangzhou 310024, China\\
$^4$International Centre for Theoretical Physics Asia-Pacific, Beijing/Hangzhou, China\\
$^5$School of Physical Sciences, University of Chinese Academy of Sciences, Beijing 100049, China
\\
$^6$MinJiang Innovative Center for Theoretical Physics, MinJiang University, Fuzhou 350108, China
}

\date{\today}

\begin{abstract}
Network dismantling aims at breaking a network into disconnected components, and attacking vertices that intersect with many loops has proven to be a most efficient strategy. But the existing loop-focusing methods treat the short loops within densely connected local clusters (e.g., cliques) as equally important as the long loops connecting different clusters. Here we propose for highly clustered artificial and real-world networks a bipartite factor-graph formulation that retains all the long loops while simplifies the local dense clusters as individual factor nodes. We develop a mean-field theory for the associated long-loop feedback vertex set problem and apply its message-passing equation as a solver for network dismantling. The proposed factor-graph loop algorithm outperforms the current state-of-the-art graph loop algorithms by a considerable margin on various real networks. Further improvement in dismantling performance is achievable by optimizing the choice of the local dense clusters.
\end{abstract}


\maketitle

Network dismantling is the optimization version of the celebrated site percolation problem. It aims at breaking a network into many disconnected components with the minimum number of vertex deletions~\cite{Het2008,Z2013,MM2015,MZ2016,Bet2016,Cet2016,Qin-2018,Ret2019}. Dismantling a network essentially means demolishing all its long-range loops and it is deeply connected to the concept of feedback vertex set (FVS, a set of vertices whose removal will break all the loops in the network~\cite{Het2008}).   The dismantling problem is rooted in many structural and dynamical issues of network science, for example contagion spreading and vaccination~\cite{Altarelli-Braunstein-DallAsta-Zecchina-2013,Cet2008}, vital vertex identification~\cite{Lv-etal-2016,Erkol-etal-2019},  control of complex systems~\cite{Liu-Barabasi-2016,Lokhov-Saad-2017}, and network robustness enhancement~\cite{Chujyo-Hayashi-2021}. Many different types of heuristic algorithms were proposed over the past years to solve this important problem~\cite{Zet2016,Zhao-Habibulla-Zhou-2015,Wet2018,Fet2020,Zet2020,WSet2020}. Among them some of the most efficient and best-performing ones (BPD~\cite{MZ2016,Qin-2018}, Min-Sum~\cite{Bet2016}, CoreHD~\cite{Zet2016,Zhao-Habibulla-Zhou-2015})  are FVS-based iterative processes that delete the vertices deemed most vital for loop integrity. The dismantling and FVS problems are equivalent for artificially generated random networks of which most loops are long-range ones~\cite{Het2008}, naturally the FVS strategy is optimal for dismantling such networks~\cite{MZ2016,Bet2016}.

Triangles and other short-range loops are however abundant in real-world networks, leading to densely connected local clusters and communities~\cite{WS1998,Girvan-Newman-2002,N2003,Palla-etal-2005,Let2009,SB2011,Fortunato-Hric-2016}, which are often desirable not to split into different parts. Deleting vertices that are highly involved in short-range loops within the dense local regions will also increase the cost of dismantling such real networks~\cite{Cet2008,WSet2020}. It should be more advantageous to delete those articulation points linking different local clusters/communities~\cite{WSet2020,Tet2017}, yet all the present FVS-based methods do not distinguish between short-range and long-range loops. 

In this Letter, we design a long-loop FVS model system and present an improved FVS approach to network dismantling. Our model represents locally dense clusters as factors in a bipartite factor-graph and intentionally neglects all the loops within these specified clusters. Many short-range loops of the original network therefore are absent in the constructed factor-graph, while all the long-range loops are preserved. We derive a coarse-grained (two-state) message-passing equation for this model to evaluate the impact of deleting a vertex to the long-range loop structure, and use this equation as a solver for dismantling artificial and real-world highly clustered networks. We find that even if only the shortest loops in cliques of three and four vertices are explicitly excluded, our factor-graph model already leads to remarkable and consistent improvements in dismantling performance as compared with the conventional network model. Our theoretical framework leaves room for the optimal design of the factor-graphs.

\begin{figure}[t]
\centering   
\includegraphics[width=1.0\linewidth]{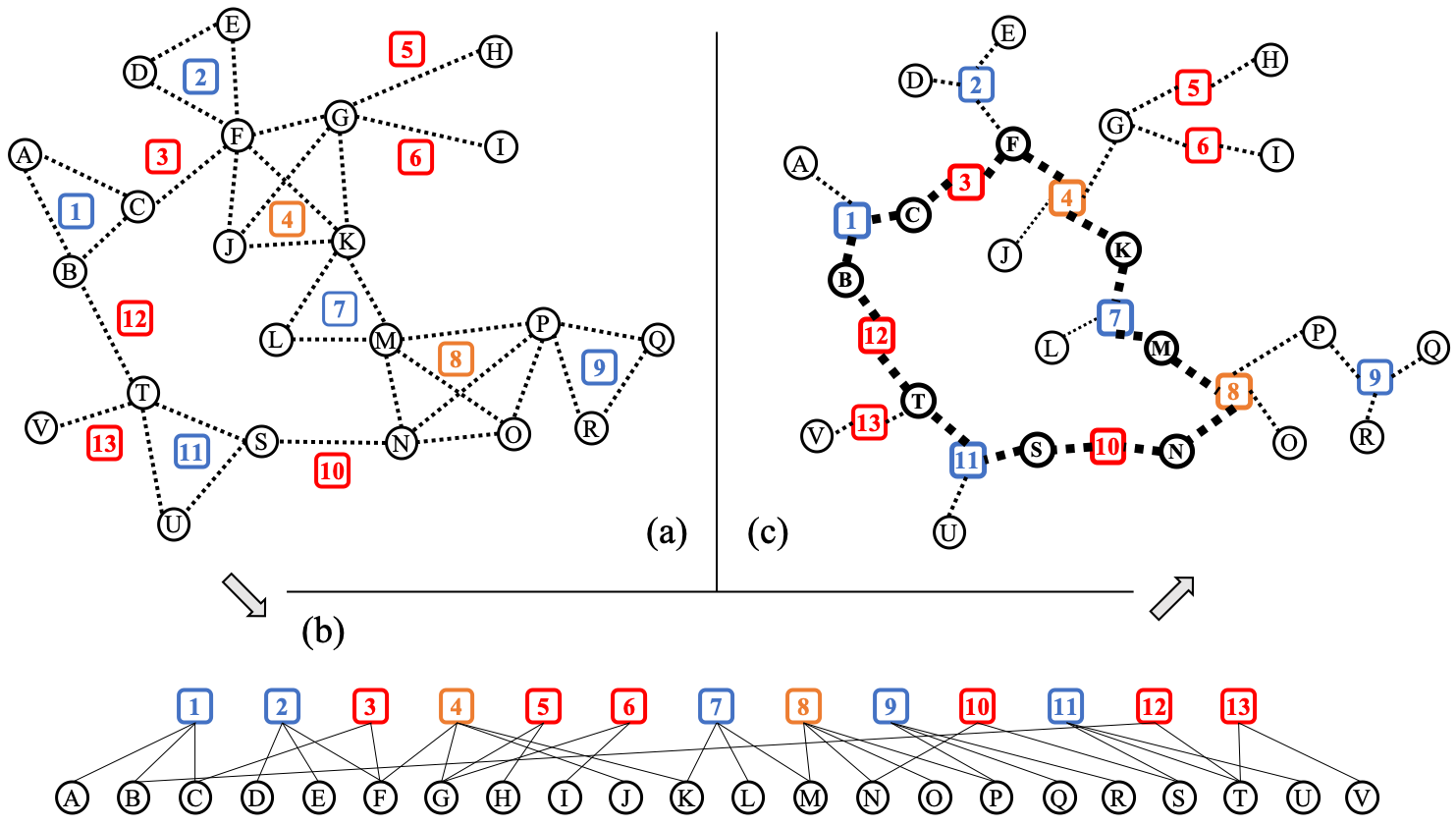}  
\caption{
Factor-graph construction for the long-loop feedback vertex set problem. This example network (a) contains $22$ vertices (circles). We specify all the maximal cliques as the local clusters and represent them as factors (squares). The resulting bipartite factor-graph (b) has $13$ factors and $22$ vertices, in which a link between a vertex and a factor indicates that the vertex is a member of the cluster represented by the factor. The loop of dotted links in (c) highlights the only inter-cluster long loop of the factor-graph.
}
\label{fig:example}
\end{figure}

\emph{Factor-graph construction.}-- Given a simple network $G$ of $N$ vertices and $M$ undirected edges between these vertices, which contains many densely connected local clusters/communities and also long-range loops, our objective is to delete a minimum number of vertices to break the long loops while preserving the local loopy structures as intact as possible. To minimize the distracting effects of the short loops to our long-loop breaking algorithms, we introduce a set of factors and expand the original network into a bipartite factor-graph, $\mathcal{G}$ (Fig.~\ref{fig:example}). Each factor (square) $a$ of $\mathcal{G}$ represents a cluster of $G$ which is a densely connected local region with a high proportion of short loops, and a link $(i, a)$ is drawn between every vertex (circle) $i$ of this cluster and the factor $a$ to indicate that $i$ is a member of cluster $a$. There is a huge literature on discovering clusters/communities in a complex network~\cite{Fortunato-Hric-2016}, and our theoretical framework is flexible to the different criteria that might be adopted in appointing the clusters. A particularly simple recipe is to specify some cliques of the graph as the clusters~\cite{Palla-etal-2005}. In our construction, if an edge of $G$ is not assigned to any such clusters, we regard the two incident vertices as forming a minimum-sized cluster (a $2$-clique). The clusters may partially overlap at some vertices and then a vertex is generally connected to many factors in $\mathcal{G}$ (Fig.~\ref{fig:example}). Notice that all the edges and loops within the individual clusters completely disappear in the constructed factor-graph $\mathcal{G}$. All the loops in $\mathcal{G}$, necessarily alternating between vertices and factors, correspond to the inter-cluster (and often long) loops of the original graph $G$.

\emph{Model with local constraints.}-- For a bipartite factor-graph $\mathcal{G}$ of vertices and factors, we define a subset $\Gamma$ of its \emph{vertices} as a feedback vertex set if it intersects with every loop of $\mathcal{G}$~\cite{Het2008}. A minimum FVS (minFVS) is then a FVS of global minimum cardinality, which offers an optimal way of breaking all the inter-cluster loops and dismantling $\mathcal{G}$ into a collection of tree components~\cite{Het2008}. However, the minFVS problem is an intrinsically hard combinatorial optimization problem and an exact solution is practically impossible~\cite{Fomin-etal-2008}. Here we derive an approximate but efficient message-passing algorithm by extending an earlier spin glass model~\cite{Z2013}. In the following discussions we use symbols $i, j, k, \ldots$ to indicate vertices and  $a, b, c, \ldots$ to indicate factors, and denote by $\partial i$ ($\partial a$) the set of nearest-neighboring factors (vertices) of vertex $i$ (factor $a$).

First, a state $c_i$ is introduced to each vertex $i$, which is either zero ($c_i \! = \! 0$, indicating $i$ being inactive) or is equal to the index $a$ of a neighboring factor ($i$ being active, $c_i \! = a \! \in \! \partial i$). A configuration of the whole system is denoted as $\underline{c} \!  \equiv \! (c_1, \ldots, c_N)$, and its energy is simply the total number of zero-state (inactive) vertices. Second, a local constraint $\chi_a$ is introduced for every factor $a$ as
\begin{equation}
\chi_a = \prod\limits_{i \in \partial a}
( \delta_{c_i}^a + \delta_{c_i}^0 )
+ \sum\limits_{i \in \partial a}
( 1 - \delta_{c_i}^a - \delta_{c_i}^0)
\prod\limits_{j\in \partial a\backslash i}
( \delta_{c_j}^a + \delta_{c_j}^0) \; ,
\label{eq:chia}
\end{equation}
where $\delta_m^n \! = \! 1$ if $m\! = \! n$ and $\delta_m^n \! = \! 0$ otherwise, and $\partial a\backslash i$ means vertex $i$ is excluded from set $\partial a$. Notice that $\chi_a \! = \! 1$  only if at most one of the neighboring vertices of factor $a$ takes state other than $0$ and $a$, otherwise $\chi_a \! = \! 0$. We require $\bm{c}$ to satisfy $\chi_a \! = \! 1$ for all the factors $a$. Given such a valid configuration $\underline{c}$, then each connected subgraph of $\mathcal{G}$ formed by the active vertices and the attached factors is most often a tree (free of any loop) or occasionally a cycle-tree (containing exactly one loop)~\cite{Z2013}. If cycle-trees do exist, we can easily break each of the associated loops by inactivate a single vertex, and then all the inactive vertices of $\underline{c}$ form a FVS. Conversely, each FVS $\Gamma$ of $\mathcal{G}$ can be mapped to a set of valid configurations $\underline{c}$ after setting $c_i \! = \! 0$ for every vertex $i \! \in \!  \Gamma$~\cite{Z2013}. 

We define the partition function $Z(\beta)$ of the system as
\begin{equation}
 Z(\beta) = \sum\limits_{\underline{c}} \prod\limits_i
 \bigl[ 1 + (e^{-\beta} - 1) \delta_{c_i}^0 \bigr] 
 \prod\limits_a \chi_a \; ,
 \label{eq:Zbeta}
\end{equation}
where $\beta$ is the inverse temperature parameter. At large values of $\beta$ the partition function will be predominantly contributed by the minFVS configurations and their low-energy excitations.

\emph{Belief-propagation (BP).}--
We solve the spin glass model (\ref{eq:Zbeta}) by the now standard replica-symmetric cavity method~\cite{Mezard-Montanari-2009,Zhou-2015}. First, the probability of vertex $i$ taking state $c_i$ in the absence of the constraint from neighboring factor $a$, denoted as $q_{i\rightarrow a}^{c_i}$, is estimated through the following self-consistent BP equation
\begin{subequations}
\label{eq:qia}
\begin{align}
q_{i\rightarrow a}^0 & \propto 
e^{-\beta} 
\prod\limits_{b\in \partial i\backslash a}
\Bigl[ 1 +
\sum\limits_{j\in \partial b\backslash i} (\frac{1}{q_{j\rightarrow b}^b
+ q_{j\rightarrow b}^0} - 1)
\Bigr]
\; ,
\\
q_{i\rightarrow a}^{b} & \propto
1+ \sum\limits_{j\in \partial b\backslash i}
(\frac{1}{q_{j\rightarrow b}^b + q_{j\rightarrow b}^0} - 1)
\; ,  \;\; (b \in \partial i\backslash a)
\\
q_{i\rightarrow a}^{a} & \propto 1 \; , 
\end{align}
\end{subequations}
with normalization $q_{i\rightarrow a}^0 \! +\! 
q_{i\rightarrow a}^a \! + \! \sum_{b\in \partial i\backslash a} q_{i\rightarrow a}^b \! =\! 1$, where $\partial i\backslash a$ means excluding factor $a$ from set $\partial i$. In the numerical implementation we need only to iterate the coarse-grained probability $q_{i\rightarrow a}^{a+0} \! \equiv \! q_{i\rightarrow a}^a \! + \! q_{i\rightarrow a}^0$, making the minFVS problem essentially an Ising-type spin glass system. Second, the marginal probability of vertex $i$ taking state $c_i \! = \! 0$ (i.e., being a feedback vertex) under the constraints of all its neighboring factors $a$, denoted as $q_i^0$, is estimated through
\begin{widetext}
\begin{equation}
q_i^0 = \frac{e^{-\beta} \prod\limits_{a\in \partial i} 
\bigl[1 + \sum\limits_{j\in \partial a\backslash i} ( \frac{1}{q_{j\rightarrow a}^a
+ q_{j\rightarrow a}^0} - 1)\bigr]}{e^{-\beta} \prod\limits_{a\in \partial i} 
\bigl[1 + \sum\limits_{j\in \partial a\backslash i} ( \frac{1}{q_{j\rightarrow a}^a
+ q_{j\rightarrow a}^0} - 1)\bigr] + \sum\limits_{a\in \partial i}
\bigl[ 1 + \sum\limits_{j\in \partial a\backslash i} (\frac{1}{q_{j\rightarrow a}^a
+q_{j\rightarrow a}^0} - 1) \bigr]} \; .
\end{equation}
\end{widetext}
The mean fraction $\rho$ of feedback vertices at inverse temperature $\beta$ is then
\begin{equation}
    \rho = \frac{1}{N} \sum\limits_{i} q_i^0 \; .
\end{equation}

Explicit expressions for the free energy density and the entropy density could also be derived and be evaluated at a BP fixed point~\cite{Let2021}. We can then estimate the minFVS relative size by taking the $\beta \! \rightarrow \! \infty$ of $\rho$ (if the entropy density is non-negative at this limit) or by taking the value of $\rho$ at the maximal value of $\beta$ at which the entropy density reaches zero from above~\cite{Z2013}.

Similar to the iterative process of the BP-guided decimation (BPD) algorithm~\cite{Z2013}, we could try to construct a close-to-minimum FVS for the factor-graph $\mathcal{G}$ by deleting at each BP iteration a tiny fraction of the vertices $i$ of $\mathcal{G}$ whose estimated inactive probability $q_i^0$ are the highest among all the remaining vertices, until all the loops of $\mathcal{G}$ are broken. We refer to this factor-graph decimation process as FBPD.

\begin{figure}[b]
\centering   
\includegraphics[width=1.0\linewidth]{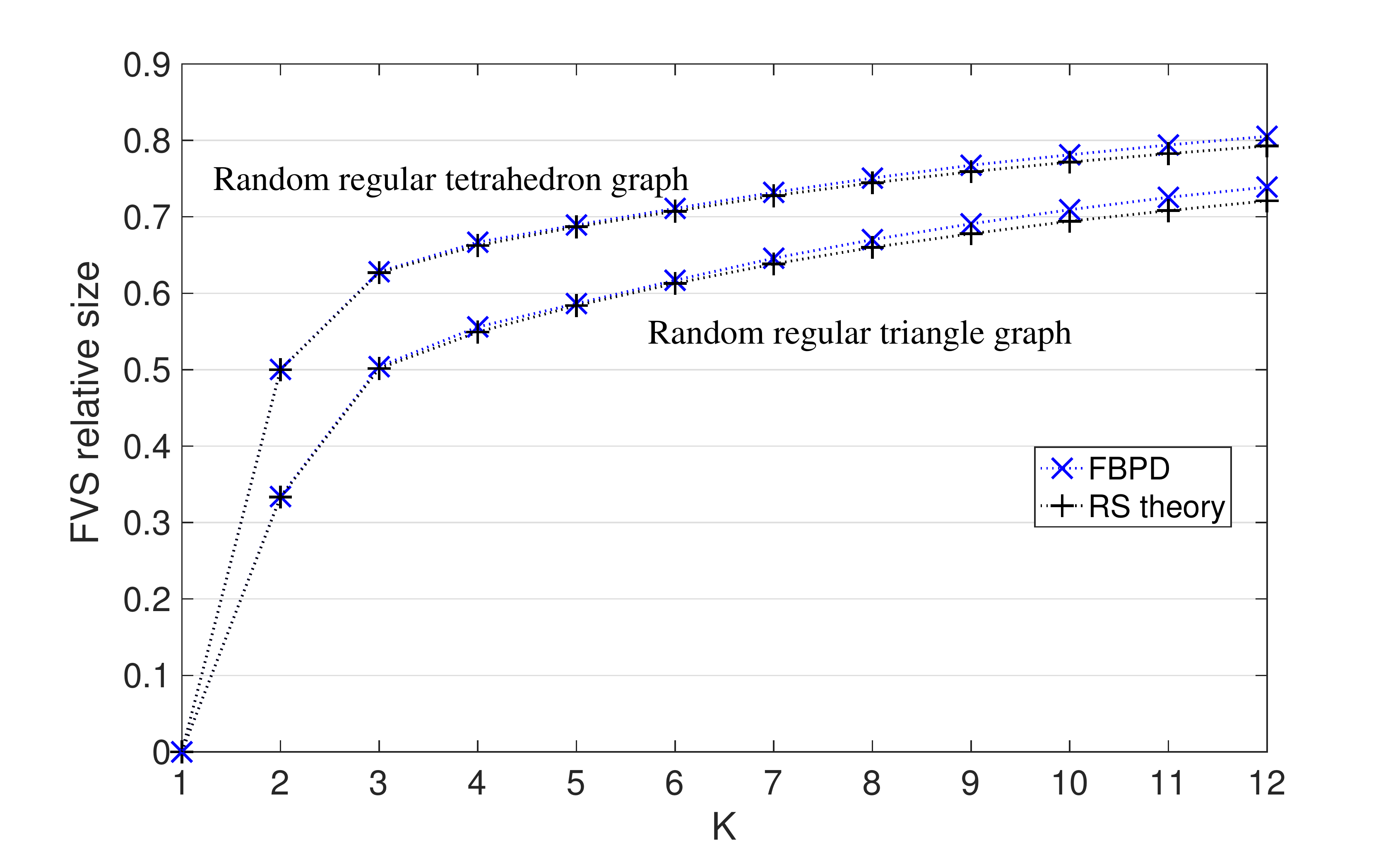}  
\caption{The minimum fraction of long-loop feedback vertices for random regular triangle networks (lower points) and tetrahedron networks (upper points). $K$ is the number of triangles/tetrahedra participated by each vertex. The FBPD algorithmic results obtained on single network instances of $N \! = \! 100002$ (crosses) are only slightly higher than the predictions of the replica-symmetric mean field theory (pluses).} 
\label{fig:rrresult}
\end{figure}

\emph{Results on random and real networks.}--
We first test the FBPD performance on random networks $G$ formed by local $n$-cliques (i.e., fully connected $n$-vertex subnetworks). Each vertex participates in exactly $K$ randomly chosen $n$-cliques and the graph is otherwise completely random~\cite{M2009, N2009,G2009,Yet2011,Z2017}. The total number of $n$-cliques is $N K / n$ while the total number of edges is $M \! = \! N K (n-1) /2$. All the vertices are involved in both intra-clique short loops and inter-clique long loops. Naturally we represent each $n$-clique as a factor, getting a factor-graph $\mathcal{G}$ that retains only the long loops and is locally tree-like. Quantitative results obtained on the cases of $n \! = \! 3$ (i.e., triangle clusters) and $n \! = \! 4$ (tetrahedron clusters) are shown in Fig.~\ref{fig:rrresult}. We find that, at a given value of $K$, the fraction $\rho$ of long-loop feedback vertices achieved by FBPD on individual network instances is only slightly exceeding the theoretical minFVS relative size $\rho_{\rm min}$ predicted by the replica-symmetric mean field theory.  For example at $K\! =\! 10$ and $n\! = \! 3$, $\rho_{\rm min} \! = \! 0.6932$ while the FBPD empirical values are $\rho\! \approx \! 0.7080$ for graphs of size $N \! = \! 10^5$ ($\beta \! = \! 7$ for FBPD, and the algorithmic results are insensitive to this parameter).

\begin{figure}[t]
\centering   
\includegraphics[width=1.0\linewidth]{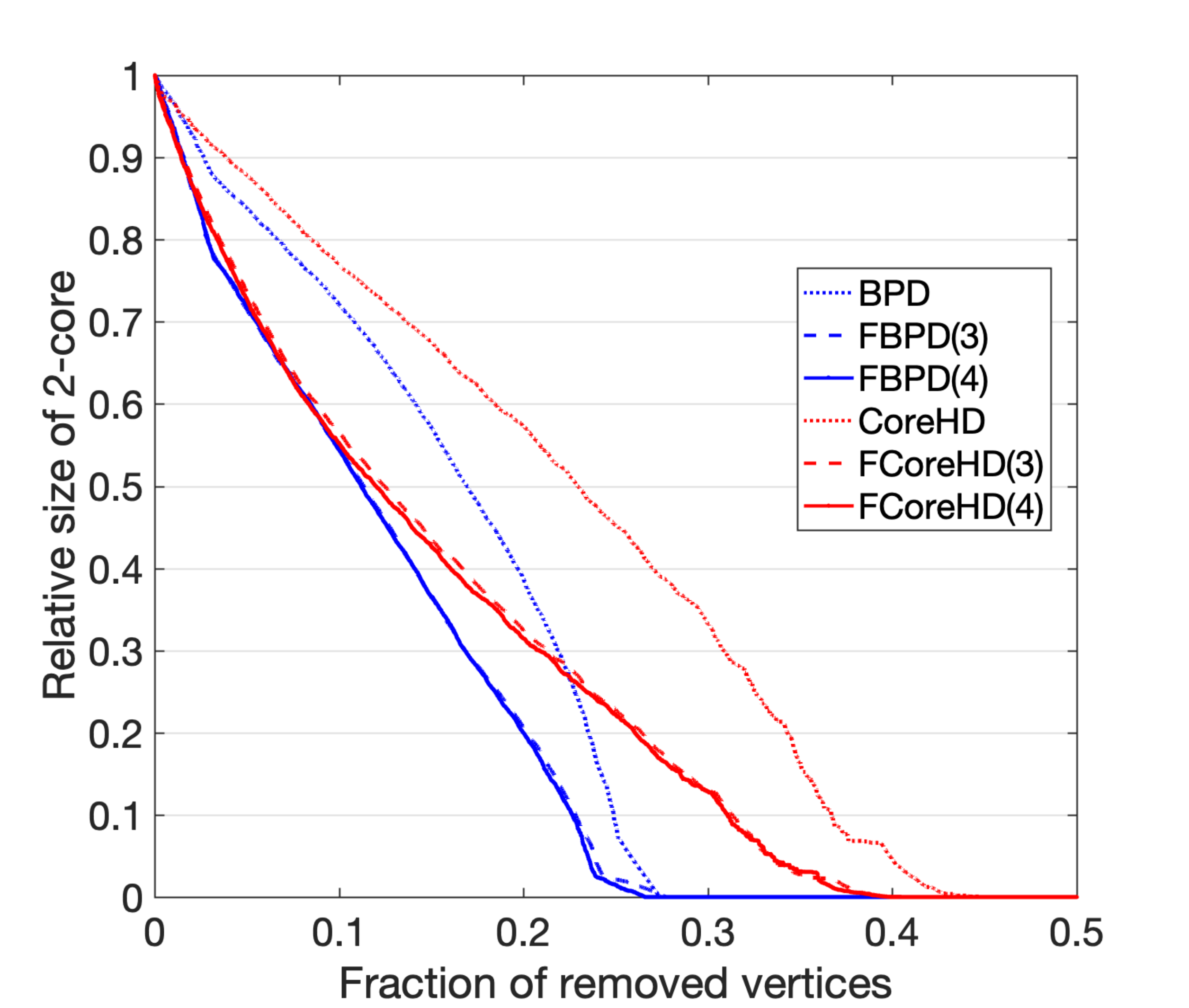}  
\caption{Shrinkage of the 2-core of the LastFM network ($7624$ vertices, $27806$ edges) as the vertices are sequentially deleted according to six different algorithms. The FBPD(n) and FCoreHD(n) trajectories are obtained on the factor-graph $\mathcal{G}_n$ considering cliques up to size $n$ ($3$ or $4$).
}
\label{fig:shrink}
\end{figure}

Besides the message-passing FBPD, the factor-graph representation is also helpful for other long-loop breaking heuristics. All the loops of network $G$ are contained in its $2$-core, the maximum subnetwork in which every vertex is connected to at least two other vertices of this subnetwork. Similarly, the $2$-core of the corresponding factor-graph $\mathcal{G}$ contains all the long loops of $G$. A simple and fast algorithm to destroy the $2$-core of $G$ is CoreHD, which recursively deletes a highest-degree vertex of the extant $2$-core~\cite{Zet2016,Zhao-Habibulla-Zhou-2015}. We have tested the factor-graph version of this algorithm (FCoreHD) on random clustered networks. FCoreHD is inferior to FBPD in performance, but it outperforms the original CoreHD algorithm. Similar comparative results are observed on real networks. The $2$-core of a real network will generally shrink much faster if its vertices are sequentially deleted according to the factor-graph algorithms (FBPD, FCoreHD) instead of the graph algorithms (BPD, CoreHD). Figure~\ref{fig:shrink} offers a concrete demonstration.

We now compare the performances of the factor-graph (FBPD, FCoreHD) and the graph (BPD, CoreHD) algorithms on the network dismantling problem. All these algorithms first delete some vertices to solve the minFVS problem and then reinsert some vertices back to the network under the constraints of every connected component containing at most $0.01 N$ vertices~\cite{MM2015,MZ2016,Bet2016,Zet2016,Wet2020}. The following real-world networks of miscellaneous sorts are used, including road network~\cite{SB2011}, power grid~\cite{WS1998}, internet (IntNet)~\cite{Let2005}, email networks (Email1/2)~\cite{Let2007a, Let2009}, protein interaction network (Yeast)~\cite{Bet2003}, product co-purchasing network (Amazon)~\cite{Let2007b}, Wikipedia network (Wiki)~\cite{Ret2019a}, citation network (Cite)~\cite{Let2005}, and social networks on various platforms (LastFM, Github, Twitch, Facebook networks FBK-a/b, Deezer, Brightkite BK)~\cite{RS2020, Ret2019a, ML2012, Ret2019b,Cet2011}. 
For each network $G$, we get two factor-graph versions ($\mathcal{G}_3$ and $\mathcal{G}_4$). $\mathcal{G}_3$ is obtained by repeatedly picking a $3$-clique at random from $G$ as a factor and then deleting all the edges of this $3$-clique; when there is no more $3$-cliques then continues with $2$-cliques (the conventional edges). $\mathcal{G}_4$ is similarly constructed but it starts with $4$-cliques and therefore more loops are regarded as intra-cluster short loops. Results listed in Table~\ref{tab:real} confirm that the factor-graph algorithms improve over the graph algorithms. We also observe that FBPD in general claims a substantial margin over the other three algorithms. In many cases, the performance of FBPD improves as more short loops are replaced by factors (e.g., using $\mathcal{G}_n$ with larger values of $n$).

\begin{table}
\caption{Dismantling results on real-world networks. $N$,  network size; $n$, number of deleted vertices by the algorithm (superscripts B, FB, C, FC indicate BPD, FBPD, CoreHD, FCoreHD, respectively; subscripts $3$ and $4$ indicate the maximum cliques are triangles and tetrahedra, respectively).}
\label{tab:real}
\begin{tabular}{c|c|c|c|c|c|c|c}
\hline
Network & $N$ & $n^{\textrm{B}}$ & $n^{\textrm{FB}}_3$ & $n^{\textrm{FB}}_4$ & $n^{\textrm{C}}$ & $n^{\textrm{FC}}_3$ & $n^{\textrm{FC}}_4$ \\
\hline
Email1    & 986  & 456 & 454 & 455 & \textbf{452} & 458 & \textbf{452} \\
Road    & 1174   &  148   & 149 & 149 & 149   & \textbf{147} & \textbf{147} \\
Yeast   & 2284 & 357 & 351 & 354 & 353  & \textbf{350} & 355 \\
FBK-a     & 4039  & 1907 & 1921 & \textbf{1872} & 1914 & 1892 & \textbf{1872}  \\
Grid     & 4941  & 312  & \textbf{301} & 304 & 317 & 302 & 305 \\
Wiki     & 5201 & 1148 & 1138 & \textbf{1134} & 1150 & 1149 & 1145 \\
IntNet     & 6474 & 160 & 159 & 159 & 160 & 161 & \textbf{156} \\
LastFM     & 7624 & 1285  &1262& \textbf{1258}& 1296  & 1296 &1276 \\
Twitch     & 9498  & 3013  & 3011 &\textbf{2992} & 3036  & 3037 & 3021  \\
FBK-b     & 13866  & 3311 & 3295 & \textbf{3293} & 3356 & 3332 & 3302  \\
Cite     & 34546  & 13433   & \textbf{13393} & 13399 & 13547 & 13484 &13475\\
Email2     & 36692  & \textbf{2616} & 2626 & 2647 & 2621 & 2648 & 2647 \\
Github     & 37700  & 6468  & \textbf{6422} & 6423 & 6554 & 6520 & 6495 \\
Deezer    & 54573  & 23805 &23751 &\textbf{23686} & 24180 & 24098 & 24112  \\
BK     & 58228  & 6225 & 6140 & \textbf{6130} & 6257 & 6213 & 6207  \\
Amazon     & 262111  & 45038  & 44862 &\textbf{44438} & 45612   & 44915 & 44803 \\
\hline
\end{tabular}%
\end{table}

\emph{Conclusion.}-- 
The factor-graph representation emphasizes the long-range loops in a complex network and allows us to rank the vertices according to their contributions to these long-range loops. Vertices that are only important for short-range loops will be ranked lowly through our factor-graph formulation. As one important application, we demonstrated that the factor-graph FVS algorithms (FBPD, FCoreHD) considerably outperformed the corresponding graph FVS algorithms (BPD, CoreHD) in dismantling real-world networks, even when only the shortest loops in $3$-cliques and $4$-cliques were excluded from entering the factor-graphs. This representation also enables us to compute the ensemble-averaged size of long-loop minimum feedback vertex sets for random clustered networks.

In the present work we only considered the simplest way of constructing factor-graphs. Optimizing the choice of the factors to achieve a nearly optimal factor-graph for a given real-wold network instance is an interesting open issue, insightful for future studies especially for practical applications. Another future direction is to extend the factor-graph framework to directed networks (whose edges have directions) and consider the more difficult problem of breaking all the long-range directed cycles.

\section*{Acknowledgement}

{T.L. thanks the Institute of Theoretical Physics of the Chinese Academy of Sciences (ITP-CAS) for hospitality. This work was supported by the National Natural Science Foundation of China Grants No.~11975295, No.~11947302, No.~11975294, and No.~12047503, and the Chinese Academy of Sciences Grant No.~QYZDJ-SSW-SYS018 and QYZDB-SSW-SYS032. Numerical simulations were carried out at the Tianwen and HPC clusters of ITP-CAS and the Tianhe-2 platform of the National Supercomputer Center in Guangzhou.}






\begin{thebibliography}{48}%
\makeatletter
\providecommand \@ifxundefined [1]{%
 \@ifx{#1\undefined}
}%
\providecommand \@ifnum [1]{%
 \ifnum #1\expandafter \@firstoftwo
 \else \expandafter \@secondoftwo
 \fi
}%
\providecommand \@ifx [1]{%
 \ifx #1\expandafter \@firstoftwo
 \else \expandafter \@secondoftwo
 \fi
}%
\providecommand \natexlab [1]{#1}%
\providecommand \enquote  [1]{``#1''}%
\providecommand \bibnamefont  [1]{#1}%
\providecommand \bibfnamefont [1]{#1}%
\providecommand \citenamefont [1]{#1}%
\providecommand \href@noop [0]{\@secondoftwo}%
\providecommand \href [0]{\begingroup \@sanitize@url \@href}%
\providecommand \@href[1]{\@@startlink{#1}\@@href}%
\providecommand \@@href[1]{\endgroup#1\@@endlink}%
\providecommand \@sanitize@url [0]{\catcode `\\12\catcode `\$12\catcode
  `\&12\catcode `\#12\catcode `\^12\catcode `\_12\catcode `\%12\relax}%
\providecommand \@@startlink[1]{}%
\providecommand \@@endlink[0]{}%
\providecommand \url  [0]{\begingroup\@sanitize@url \@url }%
\providecommand \@url [1]{\endgroup\@href {#1}{\urlprefix }}%
\providecommand \urlprefix  [0]{URL }%
\providecommand \Eprint [0]{\href }%
\providecommand \doibase [0]{http://dx.doi.org/}%
\providecommand \selectlanguage [0]{\@gobble}%
\providecommand \bibinfo  [0]{\@secondoftwo}%
\providecommand \bibfield  [0]{\@secondoftwo}%
\providecommand \translation [1]{[#1]}%
\providecommand \BibitemOpen [0]{}%
\providecommand \bibitemStop [0]{}%
\providecommand \bibitemNoStop [0]{.\EOS\space}%
\providecommand \EOS [0]{\spacefactor3000\relax}%
\providecommand \BibitemShut  [1]{\csname bibitem#1\endcsname}%
\let\auto@bib@innerbib\@empty
\bibitem [{\citenamefont {Haxell}\ \emph {et~al.}(2008)\citenamefont {Haxell},
  \citenamefont {Pikhurko},\ and\ \citenamefont {Thomason}}]{Het2008}%
  \BibitemOpen
  \bibfield  {author} {\bibinfo {author} {\bibfnamefont {P.}~\bibnamefont
  {Haxell}}, \bibinfo {author} {\bibfnamefont {O.}~\bibnamefont {Pikhurko}}, \
  and\ \bibinfo {author} {\bibfnamefont {A.}~\bibnamefont {Thomason}},\
  }\bibfield  {title} {\enquote {\bibinfo {title} {Maximum acyclic and
  fragmented sets in regular graphs},}\ }\href@noop {} {\bibfield  {journal}
  {\bibinfo  {journal} {J. Graph Theory}\ }\textbf {\bibinfo {volume} {57}},\
  \bibinfo {pages} {149--156} (\bibinfo {year} {2008})}\BibitemShut {NoStop}%
\bibitem [{\citenamefont {Zhou}(2013)}]{Z2013}%
  \BibitemOpen
  \bibfield  {author} {\bibinfo {author} {\bibfnamefont {H.-J.}\ \bibnamefont
  {Zhou}},\ }\bibfield  {title} {\enquote {\bibinfo {title} {Spin glass
  approach to the feedback vertex set problem},}\ }\href@noop {} {\bibfield
  {journal} {\bibinfo  {journal} {Eur. Phys. J. B}\ }\textbf {\bibinfo {volume}
  {86}},\ \bibinfo {pages} {455} (\bibinfo {year} {2013})}\BibitemShut
  {NoStop}%
\bibitem [{\citenamefont {Morone}\ and\ \citenamefont {Makse}(2015)}]{MM2015}%
  \BibitemOpen
  \bibfield  {author} {\bibinfo {author} {\bibfnamefont {F.}~\bibnamefont
  {Morone}}\ and\ \bibinfo {author} {\bibfnamefont {H.}~\bibnamefont {Makse}},\
  }\bibfield  {title} {\enquote {\bibinfo {title} {Influence maximization in
  complex networks through optimal percolation},}\ }\href@noop {} {\bibfield
  {journal} {\bibinfo  {journal} {Nature}\ }\textbf {\bibinfo {volume} {524}},\
  \bibinfo {pages} {65--68} (\bibinfo {year} {2015})}\BibitemShut {NoStop}%
\bibitem [{\citenamefont {Mugisha}\ and\ \citenamefont {Zhou}(2016)}]{MZ2016}%
  \BibitemOpen
  \bibfield  {author} {\bibinfo {author} {\bibfnamefont {S.}~\bibnamefont
  {Mugisha}}\ and\ \bibinfo {author} {\bibfnamefont {H.-J.}\ \bibnamefont
  {Zhou}},\ }\bibfield  {title} {\enquote {\bibinfo {title} {Identifying
  optimal targets of network attack by belief propagation},}\ }\href@noop {}
  {\bibfield  {journal} {\bibinfo  {journal} {Phys. Rev. E}\ }\textbf {\bibinfo
  {volume} {94}},\ \bibinfo {pages} {012305} (\bibinfo {year}
  {2016})}\BibitemShut {NoStop}%
\bibitem [{\citenamefont {Braunstein}\ \emph {et~al.}(2016)\citenamefont
  {Braunstein}, \citenamefont {Dall’Asta},\ and\ \citenamefont
  {Semerjian}}]{Bet2016}%
  \BibitemOpen
  \bibfield  {author} {\bibinfo {author} {\bibfnamefont {A.}~\bibnamefont
  {Braunstein}}, \bibinfo {author} {\bibfnamefont {L.}~\bibnamefont
  {Dall’Asta}}, \ and\ \bibinfo {author} {\bibfnamefont {L.}~\bibnamefont
  {Semerjian}, \bibfnamefont {G.and~Zdeborov{\'a}}},\ }\bibfield  {title}
  {\enquote {\bibinfo {title} {Network dismantling},}\ }\href@noop {}
  {\bibfield  {journal} {\bibinfo  {journal} {Proc. Natl. Acad. Sci. USA}\
  }\textbf {\bibinfo {volume} {113}},\ \bibinfo {pages} {12368--12373}
  (\bibinfo {year} {2016})}\BibitemShut {NoStop}%
\bibitem [{\citenamefont {Clusella}\ \emph {et~al.}(2016)\citenamefont
  {Clusella}, \citenamefont {Grassberger}, \citenamefont {P{\'e}rez-Reche},\
  and\ \citenamefont {Politi}}]{Cet2016}%
  \BibitemOpen
  \bibfield  {author} {\bibinfo {author} {\bibfnamefont {P.}~\bibnamefont
  {Clusella}}, \bibinfo {author} {\bibfnamefont {P.}~\bibnamefont
  {Grassberger}}, \bibinfo {author} {\bibfnamefont {F.~J.}\ \bibnamefont
  {P{\'e}rez-Reche}}, \ and\ \bibinfo {author} {\bibfnamefont {A.}~\bibnamefont
  {Politi}},\ }\bibfield  {title} {\enquote {\bibinfo {title} {Immunization and
  targeted destruction of networks using explosive percolation},}\ }\href@noop
  {} {\bibfield  {journal} {\bibinfo  {journal} {Phys. Rev. Lett.}\ }\textbf
  {\bibinfo {volume} {117}},\ \bibinfo {pages} {208301} (\bibinfo {year}
  {2016})}\BibitemShut {NoStop}%
\bibitem [{\citenamefont {Qin}(2018)}]{Qin-2018}%
  \BibitemOpen
  \bibfield  {author} {\bibinfo {author} {\bibfnamefont {S.-M.}\ \bibnamefont
  {Qin}},\ }\bibfield  {title} {\enquote {\bibinfo {title} {Spin-glass model
  for the c-dismantling problem},}\ }\href@noop {} {\bibfield  {journal}
  {\bibinfo  {journal} {Phys. Rev. E}\ }\textbf {\bibinfo {volume} {98}},\
  \bibinfo {pages} {062309} (\bibinfo {year} {2018})}\BibitemShut {NoStop}%
\bibitem [{\citenamefont {Ren}\ \emph {et~al.}(2019)\citenamefont {Ren},
  \citenamefont {Gleinig}, \citenamefont {Helbing},\ and\ \citenamefont
  {Antulov-Fantulin}}]{Ret2019}%
  \BibitemOpen
  \bibfield  {author} {\bibinfo {author} {\bibfnamefont {X.-L.}\ \bibnamefont
  {Ren}}, \bibinfo {author} {\bibfnamefont {N.}~\bibnamefont {Gleinig}},
  \bibinfo {author} {\bibfnamefont {D.}~\bibnamefont {Helbing}}, \ and\
  \bibinfo {author} {\bibfnamefont {N.}~\bibnamefont {Antulov-Fantulin}},\
  }\bibfield  {title} {\enquote {\bibinfo {title} {Generalized network
  dismantling},}\ }\href@noop {} {\bibfield  {journal} {\bibinfo  {journal}
  {Proc. Natl. Acad. Sci. USA}\ }\textbf {\bibinfo {volume} {116}},\ \bibinfo
  {pages} {6554--6559} (\bibinfo {year} {2019})}\BibitemShut {NoStop}%
\bibitem [{\citenamefont {Altarelli}\ \emph {et~al.}(2013)\citenamefont
  {Altarelli}, \citenamefont {Braunstein}, \citenamefont {Dall’Asta},\ and\
  \citenamefont {Zecchina}}]{Altarelli-Braunstein-DallAsta-Zecchina-2013}%
  \BibitemOpen
  \bibfield  {author} {\bibinfo {author} {\bibfnamefont {F.}~\bibnamefont
  {Altarelli}}, \bibinfo {author} {\bibfnamefont {A.}~\bibnamefont
  {Braunstein}}, \bibinfo {author} {\bibfnamefont {L.}~\bibnamefont
  {Dall’Asta}}, \ and\ \bibinfo {author} {\bibfnamefont {R.}~\bibnamefont
  {Zecchina}},\ }\bibfield  {title} {\enquote {\bibinfo {title} {Optimizing
  spread dynamics on graphs by message passing},}\ }\href@noop {} {\bibfield
  {journal} {\bibinfo  {journal} {J. Stat. Mech.: Theor. Exp.}\ }\textbf
  {\bibinfo {volume} {2013}},\ \bibinfo {pages} {P09011} (\bibinfo {year}
  {2013})}\BibitemShut {NoStop}%
\bibitem [{\citenamefont {Chen}\ \emph {et~al.}(2008)\citenamefont {Chen},
  \citenamefont {Paul}, \citenamefont {Havlin}, \citenamefont {Liljeros},\ and\
  \citenamefont {Stanley}}]{Cet2008}%
  \BibitemOpen
  \bibfield  {author} {\bibinfo {author} {\bibfnamefont {Y.}~\bibnamefont
  {Chen}}, \bibinfo {author} {\bibfnamefont {G.}~\bibnamefont {Paul}}, \bibinfo
  {author} {\bibfnamefont {S.}~\bibnamefont {Havlin}}, \bibinfo {author}
  {\bibfnamefont {F.}~\bibnamefont {Liljeros}}, \ and\ \bibinfo {author}
  {\bibfnamefont {H.~E.}\ \bibnamefont {Stanley}},\ }\bibfield  {title}
  {\enquote {\bibinfo {title} {Finding a better immunization strategy},}\
  }\href@noop {} {\bibfield  {journal} {\bibinfo  {journal} {Phys. Rev. Lett.}\
  }\textbf {\bibinfo {volume} {101}},\ \bibinfo {pages} {058701} (\bibinfo
  {year} {2008})}\BibitemShut {NoStop}%
\bibitem [{\citenamefont {L{\"u}}\ \emph {et~al.}(2016)\citenamefont {L{\"u}},
  \citenamefont {Chen}, \citenamefont {Ren}, \citenamefont {Zhang},
  \citenamefont {Zhang},\ and\ \citenamefont {Zhou}}]{Lv-etal-2016}%
  \BibitemOpen
  \bibfield  {author} {\bibinfo {author} {\bibfnamefont {L.}~\bibnamefont
  {L{\"u}}}, \bibinfo {author} {\bibfnamefont {D.}~\bibnamefont {Chen}},
  \bibinfo {author} {\bibfnamefont {X.-L.}\ \bibnamefont {Ren}}, \bibinfo
  {author} {\bibfnamefont {Q.-M.}\ \bibnamefont {Zhang}}, \bibinfo {author}
  {\bibfnamefont {Y.-C.}\ \bibnamefont {Zhang}}, \ and\ \bibinfo {author}
  {\bibfnamefont {T.}~\bibnamefont {Zhou}},\ }\bibfield  {title} {\enquote
  {\bibinfo {title} {Vital nodes identification in complex networks},}\
  }\href@noop {} {\bibfield  {journal} {\bibinfo  {journal} {Phys. Rep.}\
  }\textbf {\bibinfo {volume} {650}},\ \bibinfo {pages} {1--63} (\bibinfo
  {year} {2016})}\BibitemShut {NoStop}%
\bibitem [{\citenamefont {Erkol}\ \emph {et~al.}(2019)\citenamefont {Erkol},
  \citenamefont {Castellano},\ and\ \citenamefont
  {Radicchi}}]{Erkol-etal-2019}%
  \BibitemOpen
  \bibfield  {author} {\bibinfo {author} {\bibfnamefont {{\c{S}}.}~\bibnamefont
  {Erkol}}, \bibinfo {author} {\bibfnamefont {C.}~\bibnamefont {Castellano}}, \
  and\ \bibinfo {author} {\bibfnamefont {F.}~\bibnamefont {Radicchi}},\
  }\bibfield  {title} {\enquote {\bibinfo {title} {Systematic comparison
  between methods for the detection of influential spreaders in complex
  networks},}\ }\href@noop {} {\bibfield  {journal} {\bibinfo  {journal} {Sci.
  Rep.}\ }\textbf {\bibinfo {volume} {9}},\ \bibinfo {pages} {15095} (\bibinfo
  {year} {2019})}\BibitemShut {NoStop}%
\bibitem [{\citenamefont {Liu}\ and\ \citenamefont
  {Barab{\'a}si}(2016)}]{Liu-Barabasi-2016}%
  \BibitemOpen
  \bibfield  {author} {\bibinfo {author} {\bibfnamefont {Y.-Y.}\ \bibnamefont
  {Liu}}\ and\ \bibinfo {author} {\bibfnamefont {A.-L.}\ \bibnamefont
  {Barab{\'a}si}},\ }\bibfield  {title} {\enquote {\bibinfo {title} {Control
  principles of complex systems},}\ }\href@noop {} {\bibfield  {journal}
  {\bibinfo  {journal} {Rev. Mod. Phys.}\ }\textbf {\bibinfo {volume} {88}},\
  \bibinfo {pages} {035006} (\bibinfo {year} {2016})}\BibitemShut {NoStop}%
\bibitem [{\citenamefont {Lokhov}\ and\ \citenamefont
  {Saad}(2017)}]{Lokhov-Saad-2017}%
  \BibitemOpen
  \bibfield  {author} {\bibinfo {author} {\bibfnamefont {A.~Y.}\ \bibnamefont
  {Lokhov}}\ and\ \bibinfo {author} {\bibfnamefont {D.}~\bibnamefont {Saad}},\
  }\bibfield  {title} {\enquote {\bibinfo {title} {Optimal deployment of
  resources for maximizing impact in spreading processes},}\ }\href@noop {}
  {\bibfield  {journal} {\bibinfo  {journal} {Proc. Natl. Acad. Sci. USA}\
  }\textbf {\bibinfo {volume} {114}},\ \bibinfo {pages} {E8138--E8146}
  (\bibinfo {year} {2017})}\BibitemShut {NoStop}%
\bibitem [{\citenamefont {Chujyo}\ and\ \citenamefont
  {Hayashi}(2021)}]{Chujyo-Hayashi-2021}%
  \BibitemOpen
  \bibfield  {author} {\bibinfo {author} {\bibfnamefont {M.}~\bibnamefont
  {Chujyo}}\ and\ \bibinfo {author} {\bibfnamefont {Y.}~\bibnamefont
  {Hayashi}},\ }\bibfield  {title} {\enquote {\bibinfo {title} {A loop
  enhancement strategy for network robustness},}\ }\href@noop {} {\bibfield
  {journal} {\bibinfo  {journal} {Appl. Netw. Sci.}\ }\textbf {\bibinfo
  {volume} {6}},\ \bibinfo {pages} {3} (\bibinfo {year} {2021})}\BibitemShut
  {NoStop}%
\bibitem [{\citenamefont {Zdeborov{\'a}}\ \emph {et~al.}(2016)\citenamefont
  {Zdeborov{\'a}}, \citenamefont {Zhang},\ and\ \citenamefont
  {Zhou}}]{Zet2016}%
  \BibitemOpen
  \bibfield  {author} {\bibinfo {author} {\bibfnamefont {L.}~\bibnamefont
  {Zdeborov{\'a}}}, \bibinfo {author} {\bibfnamefont {P.}~\bibnamefont
  {Zhang}}, \ and\ \bibinfo {author} {\bibfnamefont {H.-J.}\ \bibnamefont
  {Zhou}},\ }\bibfield  {title} {\enquote {\bibinfo {title} {Fast and simple
  decycling and dismantling of networks},}\ }\href@noop {} {\bibfield
  {journal} {\bibinfo  {journal} {Sci. Rep.}\ }\textbf {\bibinfo {volume}
  {6}},\ \bibinfo {pages} {37954} (\bibinfo {year} {2016})}\BibitemShut
  {NoStop}%
\bibitem [{\citenamefont {Zhao}\ \emph {et~al.}(2015)\citenamefont {Zhao},
  \citenamefont {Habibulla},\ and\ \citenamefont
  {Zhou}}]{Zhao-Habibulla-Zhou-2015}%
  \BibitemOpen
  \bibfield  {author} {\bibinfo {author} {\bibfnamefont {J.-H.}\ \bibnamefont
  {Zhao}}, \bibinfo {author} {\bibfnamefont {Y.}~\bibnamefont {Habibulla}}, \
  and\ \bibinfo {author} {\bibfnamefont {H.-J.}\ \bibnamefont {Zhou}},\
  }\bibfield  {title} {\enquote {\bibinfo {title} {Statistical mechanics of the
  minimum dominating set problem},}\ }\href@noop {} {\bibfield  {journal}
  {\bibinfo  {journal} {J. Stat. Phys.}\ }\textbf {\bibinfo {volume} {159}},\
  \bibinfo {pages} {1154--1174} (\bibinfo {year} {2015})}\BibitemShut {NoStop}%
\bibitem [{\citenamefont {Wandelt}\ \emph {et~al.}(2018)\citenamefont
  {Wandelt}, \citenamefont {Sun}, \citenamefont {Feng}, \citenamefont {Zanin},\
  and\ \citenamefont {Havlin}}]{Wet2018}%
  \BibitemOpen
  \bibfield  {author} {\bibinfo {author} {\bibfnamefont {S.}~\bibnamefont
  {Wandelt}}, \bibinfo {author} {\bibfnamefont {X.}~\bibnamefont {Sun}},
  \bibinfo {author} {\bibfnamefont {D.}~\bibnamefont {Feng}}, \bibinfo {author}
  {\bibfnamefont {M.}~\bibnamefont {Zanin}}, \ and\ \bibinfo {author}
  {\bibfnamefont {S.}~\bibnamefont {Havlin}},\ }\bibfield  {title} {\enquote
  {\bibinfo {title} {A comparative analysis of approaches to
  network-dismantling},}\ }\href@noop {} {\bibfield  {journal} {\bibinfo
  {journal} {Sci. Rep.}\ }\textbf {\bibinfo {volume} {8}},\ \bibinfo {pages}
  {15315} (\bibinfo {year} {2018})}\BibitemShut {NoStop}%
\bibitem [{\citenamefont {Fan}\ \emph {et~al.}(2020)\citenamefont {Fan},
  \citenamefont {Zeng}, \citenamefont {Sun},\ and\ \citenamefont
  {Liu}}]{Fet2020}%
  \BibitemOpen
  \bibfield  {author} {\bibinfo {author} {\bibfnamefont {C.}~\bibnamefont
  {Fan}}, \bibinfo {author} {\bibfnamefont {L.}~\bibnamefont {Zeng}}, \bibinfo
  {author} {\bibfnamefont {Y.}~\bibnamefont {Sun}}, \ and\ \bibinfo {author}
  {\bibfnamefont {Y.-Y.}\ \bibnamefont {Liu}},\ }\bibfield  {title} {\enquote
  {\bibinfo {title} {Finding key players in complex networks through deep
  reinforcement learning},}\ }\href@noop {} {\bibfield  {journal} {\bibinfo
  {journal} {Nature Machine Intell.}\ }\textbf {\bibinfo {volume} {2}},\
  \bibinfo {pages} {317--324} (\bibinfo {year} {2020})}\BibitemShut {NoStop}%
\bibitem [{\citenamefont {Zhao}\ \emph {et~al.}(2020)\citenamefont {Zhao},
  \citenamefont {Yang}, \citenamefont {Han}, \citenamefont {Zhang},\ and\
  \citenamefont {Wang}}]{Zet2020}%
  \BibitemOpen
  \bibfield  {author} {\bibinfo {author} {\bibfnamefont {D.}~\bibnamefont
  {Zhao}}, \bibinfo {author} {\bibfnamefont {S.}~\bibnamefont {Yang}}, \bibinfo
  {author} {\bibfnamefont {X.}~\bibnamefont {Han}}, \bibinfo {author}
  {\bibfnamefont {S.}~\bibnamefont {Zhang}}, \ and\ \bibinfo {author}
  {\bibfnamefont {Z.}~\bibnamefont {Wang}},\ }\bibfield  {title} {\enquote
  {\bibinfo {title} {Dismantling and vertex cover of network through message
  passing},}\ }\href@noop {} {\bibfield  {journal} {\bibinfo  {journal} {IEEE
  Trans. Circ. Syst. II: Expr. Biref.}\ }\textbf {\bibinfo {volume} {67}},\
  \bibinfo {pages} {2732--2736} (\bibinfo {year} {2020})}\BibitemShut {NoStop}%
\bibitem [{\citenamefont {Wandelt}\ \emph {et~al.}(2020)\citenamefont
  {Wandelt}, \citenamefont {Shi}, \citenamefont {Sun},\ and\ \citenamefont
  {Zanin}}]{WSet2020}%
  \BibitemOpen
  \bibfield  {author} {\bibinfo {author} {\bibfnamefont {S.}~\bibnamefont
  {Wandelt}}, \bibinfo {author} {\bibfnamefont {X.}~\bibnamefont {Shi}},
  \bibinfo {author} {\bibfnamefont {X.}~\bibnamefont {Sun}}, \ and\ \bibinfo
  {author} {\bibfnamefont {M.}~\bibnamefont {Zanin}},\ }\bibfield  {title}
  {\enquote {\bibinfo {title} {Community detection boosts network dismantling
  on real-world networks},}\ }\href@noop {} {\bibfield  {journal} {\bibinfo
  {journal} {IEEE Access}\ }\textbf {\bibinfo {volume} {8}},\ \bibinfo {pages}
  {111954--111965} (\bibinfo {year} {2020})}\BibitemShut {NoStop}%
\bibitem [{\citenamefont {Watts}\ and\ \citenamefont
  {Strogatz}(1998)}]{WS1998}%
  \BibitemOpen
  \bibfield  {author} {\bibinfo {author} {\bibfnamefont {D.~J.}\ \bibnamefont
  {Watts}}\ and\ \bibinfo {author} {\bibfnamefont {S.~H.}\ \bibnamefont
  {Strogatz}},\ }\bibfield  {title} {\enquote {\bibinfo {title} {Collective
  dynamics of `small-world' networks},}\ }\href@noop {} {\bibfield  {journal}
  {\bibinfo  {journal} {nature}\ }\textbf {\bibinfo {volume} {393}},\ \bibinfo
  {pages} {440--442} (\bibinfo {year} {1998})}\BibitemShut {NoStop}%
\bibitem [{\citenamefont {Girvan}\ and\ \citenamefont
  {Newman}(2002)}]{Girvan-Newman-2002}%
  \BibitemOpen
  \bibfield  {author} {\bibinfo {author} {\bibfnamefont {M.}~\bibnamefont
  {Girvan}}\ and\ \bibinfo {author} {\bibfnamefont {M.~E.~J.}\ \bibnamefont
  {Newman}},\ }\bibfield  {title} {\enquote {\bibinfo {title} {Community
  structure in social and biological networks},}\ }\href@noop {} {\bibfield
  {journal} {\bibinfo  {journal} {Proc. Natl. Acad. Sci. USA}\ }\textbf
  {\bibinfo {volume} {99}},\ \bibinfo {pages} {7821--7826} (\bibinfo {year}
  {2002})}\BibitemShut {NoStop}%
\bibitem [{\citenamefont {Newman}(2003)}]{N2003}%
  \BibitemOpen
  \bibfield  {author} {\bibinfo {author} {\bibfnamefont {M.~E.~J.}\
  \bibnamefont {Newman}},\ }\bibfield  {title} {\enquote {\bibinfo {title}
  {Properties of highly clustered networks},}\ }\href@noop {} {\bibfield
  {journal} {\bibinfo  {journal} {Phys. Rev. E}\ }\textbf {\bibinfo {volume}
  {68}},\ \bibinfo {pages} {026121} (\bibinfo {year} {2003})}\BibitemShut
  {NoStop}%
\bibitem [{\citenamefont {Palla}\ \emph {et~al.}(2005)\citenamefont {Palla},
  \citenamefont {Der{\'e}nyi}, \citenamefont {Farkas},\ and\ \citenamefont
  {Vicsek}}]{Palla-etal-2005}%
  \BibitemOpen
  \bibfield  {author} {\bibinfo {author} {\bibfnamefont {G.}~\bibnamefont
  {Palla}}, \bibinfo {author} {\bibfnamefont {I.}~\bibnamefont {Der{\'e}nyi}},
  \bibinfo {author} {\bibfnamefont {I.}~\bibnamefont {Farkas}}, \ and\ \bibinfo
  {author} {\bibfnamefont {T.}~\bibnamefont {Vicsek}},\ }\bibfield  {title}
  {\enquote {\bibinfo {title} {Uncovering the overlapping community structure
  of complex networks in nature and society},}\ }\href@noop {} {\bibfield
  {journal} {\bibinfo  {journal} {Nature}\ }\textbf {\bibinfo {volume} {435}},\
  \bibinfo {pages} {814--818} (\bibinfo {year} {2005})}\BibitemShut {NoStop}%
\bibitem [{\citenamefont {Leskovec}\ \emph {et~al.}(2009)\citenamefont
  {Leskovec}, \citenamefont {Lang}, \citenamefont {Dasgupta},\ and\
  \citenamefont {Mahoney}}]{Let2009}%
  \BibitemOpen
  \bibfield  {author} {\bibinfo {author} {\bibfnamefont {J.}~\bibnamefont
  {Leskovec}}, \bibinfo {author} {\bibfnamefont {K.~J.}\ \bibnamefont {Lang}},
  \bibinfo {author} {\bibfnamefont {A.}~\bibnamefont {Dasgupta}}, \ and\
  \bibinfo {author} {\bibfnamefont {M.~W.}\ \bibnamefont {Mahoney}},\
  }\bibfield  {title} {\enquote {\bibinfo {title} {Community structure in large
  networks: Natural cluster sizes and the absence of large well-defined
  clusters},}\ }\href@noop {} {\bibfield  {journal} {\bibinfo  {journal}
  {Internet Mathematics}\ }\textbf {\bibinfo {volume} {6}},\ \bibinfo {pages}
  {29--123} (\bibinfo {year} {2009})}\BibitemShut {NoStop}%
\bibitem [{\citenamefont {{\v{S}}ubelj}\ and\ \citenamefont
  {Bajec}(2011)}]{SB2011}%
  \BibitemOpen
  \bibfield  {author} {\bibinfo {author} {\bibfnamefont {L.}~\bibnamefont
  {{\v{S}}ubelj}}\ and\ \bibinfo {author} {\bibfnamefont {M.}~\bibnamefont
  {Bajec}},\ }\bibfield  {title} {\enquote {\bibinfo {title} {Robust network
  community detection using balanced propagation},}\ }\href@noop {} {\bibfield
  {journal} {\bibinfo  {journal} {Eur. Phys. J. B}\ }\textbf {\bibinfo {volume}
  {81}},\ \bibinfo {pages} {353--362} (\bibinfo {year} {2011})}\BibitemShut
  {NoStop}%
\bibitem [{\citenamefont {Fortunato}\ and\ \citenamefont
  {Hric}(2016)}]{Fortunato-Hric-2016}%
  \BibitemOpen
  \bibfield  {author} {\bibinfo {author} {\bibfnamefont {S.}~\bibnamefont
  {Fortunato}}\ and\ \bibinfo {author} {\bibfnamefont {D.}~\bibnamefont
  {Hric}},\ }\bibfield  {title} {\enquote {\bibinfo {title} {Community
  detection in networks: A user guide},}\ }\href@noop {} {\bibfield  {journal}
  {\bibinfo  {journal} {Phys. Rep.}\ }\textbf {\bibinfo {volume} {659}},\
  \bibinfo {pages} {1--44} (\bibinfo {year} {2016})}\BibitemShut {NoStop}%
\bibitem [{\citenamefont {Tian}\ \emph {et~al.}(2017)\citenamefont {Tian},
  \citenamefont {Bashan}, \citenamefont {Shi},\ and\ \citenamefont
  {Liu}}]{Tet2017}%
  \BibitemOpen
  \bibfield  {author} {\bibinfo {author} {\bibfnamefont {L.}~\bibnamefont
  {Tian}}, \bibinfo {author} {\bibfnamefont {A.}~\bibnamefont {Bashan}},
  \bibinfo {author} {\bibfnamefont {D.-N.}\ \bibnamefont {Shi}}, \ and\
  \bibinfo {author} {\bibfnamefont {Y.-Y.}\ \bibnamefont {Liu}},\ }\bibfield
  {title} {\enquote {\bibinfo {title} {Articulation points in complex
  networks},}\ }\href@noop {} {\bibfield  {journal} {\bibinfo  {journal}
  {Nature Commun.}\ }\textbf {\bibinfo {volume} {8}},\ \bibinfo {pages} {14223}
  (\bibinfo {year} {2017})}\BibitemShut {NoStop}%
\bibitem [{\citenamefont {Fomin}\ \emph {et~al.}(2008)\citenamefont {Fomin},
  \citenamefont {Gaspers}, \citenamefont {Pyatkin},\ and\ \citenamefont
  {Razgon}}]{Fomin-etal-2008}%
  \BibitemOpen
  \bibfield  {author} {\bibinfo {author} {\bibfnamefont {F.~V.}\ \bibnamefont
  {Fomin}}, \bibinfo {author} {\bibfnamefont {S.}~\bibnamefont {Gaspers}},
  \bibinfo {author} {\bibfnamefont {A.~V.}\ \bibnamefont {Pyatkin}}, \ and\
  \bibinfo {author} {\bibfnamefont {I.}~\bibnamefont {Razgon}},\ }\bibfield
  {title} {\enquote {\bibinfo {title} {On the minimum feedback vertex set
  problem: Exact and enumeration algorithms},}\ }\href@noop {} {\bibfield
  {journal} {\bibinfo  {journal} {Algorithmica}\ }\textbf {\bibinfo {volume}
  {52}},\ \bibinfo {pages} {293--307} (\bibinfo {year} {2008})}\BibitemShut
  {NoStop}%
\bibitem [{\citenamefont {M\'ezard}\ and\ \citenamefont
  {Montanari}(2009)}]{Mezard-Montanari-2009}%
  \BibitemOpen
  \bibfield  {author} {\bibinfo {author} {\bibfnamefont {M.}~\bibnamefont
  {M\'ezard}}\ and\ \bibinfo {author} {\bibfnamefont {A.}~\bibnamefont
  {Montanari}},\ }\href@noop {} {\emph {\bibinfo {title} {Information, Physics,
  and Computation}}}\ (\bibinfo  {publisher} {Oxford University Press},\
  \bibinfo {address} {Oxford, UK},\ \bibinfo {year} {2009})\BibitemShut
  {NoStop}%
\bibitem [{\citenamefont {Zhou}(2015)}]{Zhou-2015}%
  \BibitemOpen
  \bibfield  {author} {\bibinfo {author} {\bibfnamefont {H.-J.}\ \bibnamefont
  {Zhou}},\ }\href@noop {} {\emph {\bibinfo {title} {Spin Glass and Message
  Passing}}}\ (\bibinfo  {publisher} {Science Press},\ \bibinfo {address}
  {Beijing, China},\ \bibinfo {year} {2015})\BibitemShut {NoStop}%
\bibitem [{\citenamefont {Li}\ \emph {et~al.}(2021)\citenamefont {Li},
  \citenamefont {Zhang},\ and\ \citenamefont {Zhou}}]{Let2021}%
  \BibitemOpen
  \bibfield  {author} {\bibinfo {author} {\bibfnamefont {T.}~\bibnamefont
  {Li}}, \bibinfo {author} {\bibfnamefont {P.}~\bibnamefont {Zhang}}, \ and\
  \bibinfo {author} {\bibfnamefont {H.-J.}\ \bibnamefont {Zhou}},\ }\href@noop
  {} {\bibfield  {journal} {\bibinfo  {journal} {Unpublished manuscript}\ }
  (\bibinfo {year} {2021})}\BibitemShut {NoStop}%
\bibitem [{\citenamefont {Miller}(2009)}]{M2009}%
  \BibitemOpen
  \bibfield  {author} {\bibinfo {author} {\bibfnamefont {J.~C.}\ \bibnamefont
  {Miller}},\ }\bibfield  {title} {\enquote {\bibinfo {title} {Percolation and
  epidemics in random clustered networks},}\ }\href@noop {} {\bibfield
  {journal} {\bibinfo  {journal} {Phys. Rev. E}\ }\textbf {\bibinfo {volume}
  {80}},\ \bibinfo {pages} {020901(R)} (\bibinfo {year} {2009})}\BibitemShut
  {NoStop}%
\bibitem [{\citenamefont {Newman}(2009)}]{N2009}%
  \BibitemOpen
  \bibfield  {author} {\bibinfo {author} {\bibfnamefont {M.~E.~J.}\
  \bibnamefont {Newman}},\ }\bibfield  {title} {\enquote {\bibinfo {title}
  {Random graphs with clustering},}\ }\href@noop {} {\bibfield  {journal}
  {\bibinfo  {journal} {Phys. Rev. Lett.}\ }\textbf {\bibinfo {volume} {103}},\
  \bibinfo {pages} {058701} (\bibinfo {year} {2009})}\BibitemShut {NoStop}%
\bibitem [{\citenamefont {Gleeson}(2009)}]{G2009}%
  \BibitemOpen
  \bibfield  {author} {\bibinfo {author} {\bibfnamefont {J.~P.}\ \bibnamefont
  {Gleeson}},\ }\bibfield  {title} {\enquote {\bibinfo {title} {Bond
  percolation on a class of clustered random networks},}\ }\href@noop {}
  {\bibfield  {journal} {\bibinfo  {journal} {Phys. Rev. E}\ }\textbf {\bibinfo
  {volume} {80}},\ \bibinfo {pages} {036107} (\bibinfo {year}
  {2009})}\BibitemShut {NoStop}%
\bibitem [{\citenamefont {Yoon}\ \emph {et~al.}(2011)\citenamefont {Yoon},
  \citenamefont {Goltsev}, \citenamefont {Dorogovtsev},\ and\ \citenamefont
  {Mendes}}]{Yet2011}%
  \BibitemOpen
  \bibfield  {author} {\bibinfo {author} {\bibfnamefont {S.}~\bibnamefont
  {Yoon}}, \bibinfo {author} {\bibfnamefont {A.~V.}\ \bibnamefont {Goltsev}},
  \bibinfo {author} {\bibfnamefont {S.~N.}\ \bibnamefont {Dorogovtsev}}, \ and\
  \bibinfo {author} {\bibfnamefont {J.~F.~F.}\ \bibnamefont {Mendes}},\
  }\bibfield  {title} {\enquote {\bibinfo {title} {Belief-propagation algorithm
  and the ising model on networks with arbitrary distributions of motifs},}\
  }\href@noop {} {\bibfield  {journal} {\bibinfo  {journal} {Phys. Rev. E}\
  }\textbf {\bibinfo {volume} {84}},\ \bibinfo {pages} {041144} (\bibinfo
  {year} {2011})}\BibitemShut {NoStop}%
\bibitem [{\citenamefont {Zhang}(2017)}]{Z2017}%
  \BibitemOpen
  \bibfield  {author} {\bibinfo {author} {\bibfnamefont {P.}~\bibnamefont
  {Zhang}},\ }\bibfield  {title} {\enquote {\bibinfo {title} {Spectral
  estimation of the percolation transition in clustered networks},}\
  }\href@noop {} {\bibfield  {journal} {\bibinfo  {journal} {Phys. Rev. E}\
  }\textbf {\bibinfo {volume} {96}},\ \bibinfo {pages} {042303} (\bibinfo
  {year} {2017})}\BibitemShut {NoStop}%
\bibitem [{\citenamefont {Wang}\ \emph {et~al.}(2020)\citenamefont {Wang},
  \citenamefont {Sun}, \citenamefont {Yuan}, \citenamefont {Rui},\ and\
  \citenamefont {Yang}}]{Wet2020}%
  \BibitemOpen
  \bibfield  {author} {\bibinfo {author} {\bibfnamefont {Z.}~\bibnamefont
  {Wang}}, \bibinfo {author} {\bibfnamefont {C.}~\bibnamefont {Sun}}, \bibinfo
  {author} {\bibfnamefont {G.}~\bibnamefont {Yuan}}, \bibinfo {author}
  {\bibfnamefont {X.}~\bibnamefont {Rui}}, \ and\ \bibinfo {author}
  {\bibfnamefont {X.}~\bibnamefont {Yang}},\ }\bibfield  {title} {\enquote
  {\bibinfo {title} {A neighborhood link sensitive dismantling method for
  social networks},}\ }\href@noop {} {\bibfield  {journal} {\bibinfo  {journal}
  {J. Comput. Sci.}\ }\textbf {\bibinfo {volume} {43}},\ \bibinfo {pages}
  {101129} (\bibinfo {year} {2020})}\BibitemShut {NoStop}%
\bibitem [{\citenamefont {Leskovec}\ \emph {et~al.}(2005)\citenamefont
  {Leskovec}, \citenamefont {Kleinberg},\ and\ \citenamefont
  {Faloutsos}}]{Let2005}%
  \BibitemOpen
  \bibfield  {author} {\bibinfo {author} {\bibfnamefont {J.}~\bibnamefont
  {Leskovec}}, \bibinfo {author} {\bibfnamefont {J.}~\bibnamefont {Kleinberg}},
  \ and\ \bibinfo {author} {\bibfnamefont {C.}~\bibnamefont {Faloutsos}},\
  }\bibfield  {title} {\enquote {\bibinfo {title} {Graphs over time:
  densification laws, shrinking diameters and possible explanations},}\ }in\
  \href@noop {} {\emph {\bibinfo {booktitle} {Proc. 11st ACM SIGKDD Int. Conf.
  Knowledge Discovery in Data Mining}}}\ (\bibinfo {year} {2005})\ pp.\
  \bibinfo {pages} {177--187}\BibitemShut {NoStop}%
\bibitem [{\citenamefont {Leskovec}\ \emph
  {et~al.}(2007{\natexlab{a}})\citenamefont {Leskovec}, \citenamefont
  {Kleinberg},\ and\ \citenamefont {Faloutsos}}]{Let2007a}%
  \BibitemOpen
  \bibfield  {author} {\bibinfo {author} {\bibfnamefont {J.}~\bibnamefont
  {Leskovec}}, \bibinfo {author} {\bibfnamefont {J.}~\bibnamefont {Kleinberg}},
  \ and\ \bibinfo {author} {\bibfnamefont {C.}~\bibnamefont {Faloutsos}},\
  }\bibfield  {title} {\enquote {\bibinfo {title} {Graph evolution:
  Densification and shrinking diameters},}\ }\href@noop {} {\bibfield
  {journal} {\bibinfo  {journal} {ACM Trans. Knowledge Discovery from Data
  (TKDD)}\ }\textbf {\bibinfo {volume} {1}},\ \bibinfo {pages} {2--es}
  (\bibinfo {year} {2007}{\natexlab{a}})}\BibitemShut {NoStop}%
\bibitem [{\citenamefont {Bu}\ \emph {et~al.}(2003)\citenamefont {Bu},
  \citenamefont {Zhao}, \citenamefont {Cai}, \citenamefont {Xue}, \citenamefont
  {Zhu}, \citenamefont {Lu}, \citenamefont {Zhang}, \citenamefont {Sun},
  \citenamefont {Ling}, \citenamefont {Zhang}, \citenamefont {Li},\ and\
  \citenamefont {Chen}}]{Bet2003}%
  \BibitemOpen
  \bibfield  {author} {\bibinfo {author} {\bibfnamefont {D.}~\bibnamefont
  {Bu}}, \bibinfo {author} {\bibfnamefont {Y.}~\bibnamefont {Zhao}}, \bibinfo
  {author} {\bibfnamefont {L.}~\bibnamefont {Cai}}, \bibinfo {author}
  {\bibfnamefont {H.}~\bibnamefont {Xue}}, \bibinfo {author} {\bibfnamefont
  {X.}~\bibnamefont {Zhu}}, \bibinfo {author} {\bibfnamefont {H.}~\bibnamefont
  {Lu}}, \bibinfo {author} {\bibfnamefont {J.}~\bibnamefont {Zhang}}, \bibinfo
  {author} {\bibfnamefont {S.}~\bibnamefont {Sun}}, \bibinfo {author}
  {\bibfnamefont {L.}~\bibnamefont {Ling}}, \bibinfo {author} {\bibfnamefont
  {N.}~\bibnamefont {Zhang}}, \bibinfo {author} {\bibfnamefont
  {G.}~\bibnamefont {Li}}, \ and\ \bibinfo {author} {\bibfnamefont
  {R.}~\bibnamefont {Chen}},\ }\bibfield  {title} {\enquote {\bibinfo {title}
  {Topological structure analysis of the protein--protein interaction network
  in budding yeast},}\ }\href@noop {} {\bibfield  {journal} {\bibinfo
  {journal} {Nucleic Acids Res.}\ }\textbf {\bibinfo {volume} {31}},\ \bibinfo
  {pages} {2443--2450} (\bibinfo {year} {2003})}\BibitemShut {NoStop}%
\bibitem [{\citenamefont {Leskovec}\ \emph
  {et~al.}(2007{\natexlab{b}})\citenamefont {Leskovec}, \citenamefont
  {Adamic},\ and\ \citenamefont {Huberman}}]{Let2007b}%
  \BibitemOpen
  \bibfield  {author} {\bibinfo {author} {\bibfnamefont {J.}~\bibnamefont
  {Leskovec}}, \bibinfo {author} {\bibfnamefont {L.~A.}\ \bibnamefont
  {Adamic}}, \ and\ \bibinfo {author} {\bibfnamefont {B.~A.}\ \bibnamefont
  {Huberman}},\ }\bibfield  {title} {\enquote {\bibinfo {title} {The dynamics
  of viral marketing},}\ }\href@noop {} {\bibfield  {journal} {\bibinfo
  {journal} {ACM Trans. Web (TWEB)}\ }\textbf {\bibinfo {volume} {1}},\
  \bibinfo {pages} {5--es} (\bibinfo {year} {2007}{\natexlab{b}})}\BibitemShut
  {NoStop}%
\bibitem [{\citenamefont {Rozemberczki}\ \emph
  {et~al.}(2019{\natexlab{a}})\citenamefont {Rozemberczki}, \citenamefont
  {Allen},\ and\ \citenamefont {Sarkar}}]{Ret2019a}%
  \BibitemOpen
  \bibfield  {author} {\bibinfo {author} {\bibfnamefont {B.}~\bibnamefont
  {Rozemberczki}}, \bibinfo {author} {\bibfnamefont {C.}~\bibnamefont {Allen}},
  \ and\ \bibinfo {author} {\bibfnamefont {R.}~\bibnamefont {Sarkar}},\
  }\bibfield  {title} {\enquote {\bibinfo {title} {Multi-scale attributed node
  embedding},}\ }\href@noop {} {\bibfield  {journal} {\bibinfo  {journal}
  {arXiv preprint arXiv:1909.13021}\ } (\bibinfo {year}
  {2019}{\natexlab{a}})}\BibitemShut {NoStop}%
\bibitem [{\citenamefont {Rozemberczki}\ and\ \citenamefont
  {Sarkar}(2020)}]{RS2020}%
  \BibitemOpen
  \bibfield  {author} {\bibinfo {author} {\bibfnamefont {B.}~\bibnamefont
  {Rozemberczki}}\ and\ \bibinfo {author} {\bibfnamefont {R.}~\bibnamefont
  {Sarkar}},\ }\bibfield  {title} {\enquote {\bibinfo {title} {Characteristic
  functions on graphs: Birds of a feather, from statistical descriptors to
  parametric models},}\ }in\ \href@noop {} {\emph {\bibinfo {booktitle} {Proc.
  29th ACM Int. Conf. Information and Knowledge Management}}}\ (\bibinfo {year}
  {2020})\ pp.\ \bibinfo {pages} {1325--1334}\BibitemShut {NoStop}%
\bibitem [{\citenamefont {McAuley}\ and\ \citenamefont
  {Leskovec}(2012)}]{ML2012}%
  \BibitemOpen
  \bibfield  {author} {\bibinfo {author} {\bibfnamefont {J.~J.}\ \bibnamefont
  {McAuley}}\ and\ \bibinfo {author} {\bibfnamefont {J.}~\bibnamefont
  {Leskovec}},\ }\bibfield  {title} {\enquote {\bibinfo {title} {Learning to
  discover social circles in ego networks.}}\ }\href@noop {} {\bibfield
  {journal} {\bibinfo  {journal} {Adv. Neur. Inf. Proc. Syst.}\ }\textbf
  {\bibinfo {volume} {2012}},\ \bibinfo {pages} {548--556} (\bibinfo {year}
  {2012})}\BibitemShut {NoStop}%
\bibitem [{\citenamefont {Rozemberczki}\ \emph
  {et~al.}(2019{\natexlab{b}})\citenamefont {Rozemberczki}, \citenamefont
  {Davies}, \citenamefont {Sarkar},\ and\ \citenamefont {Sutton}}]{Ret2019b}%
  \BibitemOpen
  \bibfield  {author} {\bibinfo {author} {\bibfnamefont {B.}~\bibnamefont
  {Rozemberczki}}, \bibinfo {author} {\bibfnamefont {R.}~\bibnamefont
  {Davies}}, \bibinfo {author} {\bibfnamefont {R.}~\bibnamefont {Sarkar}}, \
  and\ \bibinfo {author} {\bibfnamefont {C.}~\bibnamefont {Sutton}},\
  }\bibfield  {title} {\enquote {\bibinfo {title} {Gemsec: Graph embedding with
  self clustering},}\ }in\ \href@noop {} {\emph {\bibinfo {booktitle} {Proc.
  2019 IEEE/ACM Int. Conf. Advances in Social Networks Analysis and Mining}}}\
  (\bibinfo {year} {2019})\ pp.\ \bibinfo {pages} {65--72}\BibitemShut
  {NoStop}%
\bibitem [{\citenamefont {Cho}\ and\ \citenamefont {Leskovec}(2011)}]{Cet2011}%
  \BibitemOpen
  \bibfield  {author} {\bibinfo {author} {\bibfnamefont {S.~A.}\ \bibnamefont
  {Cho}, \bibfnamefont {E.and~Myers}}\ and\ \bibinfo {author} {\bibfnamefont
  {J.}~\bibnamefont {Leskovec}},\ }\bibfield  {title} {\enquote {\bibinfo
  {title} {Friendship and mobility: user movement in location-based social
  networks},}\ }in\ \href@noop {} {\emph {\bibinfo {booktitle} {Proc. 17th ACM
  SIGKDD Int. Conf. Knowledge Discovery and Data Mining}}}\ (\bibinfo {year}
  {2011})\ pp.\ \bibinfo {pages} {1082--1090}\BibitemShut {NoStop}%
\end{thebibliography}

%

\end{document}